\documentclass[twocolumn,pra,aps,superscriptaddress,showpacs]{revtex4-1}

 \usepackage{mathptmx}
\usepackage{dcolumn}
\usepackage{amsmath,amssymb,amsfonts}
\usepackage{dsfont}
\usepackage{bm}
\usepackage{color}
\usepackage{latexsym}
\usepackage{epstopdf}
\usepackage[english]{babel}
\usepackage{enumerate}

\usepackage[pdftex]{graphicx}

\usepackage{natbib}
\usepackage{epstopdf}
\usepackage{appendix}

\definecolor{mygrey}{gray}{0.35}
\definecolor{myblue}{rgb}{0.2,0.2,0.8}
\definecolor{myzard}{cmyk}{0,0,0.05,0}
\definecolor{mywhite}{rgb}{1,1,1}
\definecolor{mywhite}{rgb}{1,1,1}
\definecolor{myred}{rgb}{1,0.,0.3}

\usepackage[colorlinks=true,citecolor=myblue,linkcolor=myred]{hyperref}

\newcommand{\ket}[1]{\left| #1\right\rangle}

\newcommand\nn{\mathbf{n}}
\newcommand\mm{\mathbf{m}}

\newcommand\kk{\mathbf{k}}

\newcommand\EE{\mathbf{E}}

\newcommand\RR{\mathbf{R}}

\newcommand\intt{\mathrm{int}}

\begin{document}

\title{Cold atoms in twisted bilayer optical potentials}
 \author{A.~Gonz\'{a}lez-Tudela}
 \email{a.gonzalez.tudela@csic.es}
 \affiliation{Instituto de F\'isica Fundamental IFF-CSIC, Calle Serrano 113b, Madrid 28006, Spain.}
   \author{J. I.~Cirac}
  \affiliation{Max-Planck-Institut f\"{u}r Quantenoptik Hans-Kopfermann-Str. 1. 85748 Garching, Germany }
\begin{abstract}
	The possibility of creating crystal bilayers twisted with respect to each other has led to the discovery of a wide range of novel electron correlated phenomena whose full understanding is still under debate. Here we propose and analyze a method to simulate twisted bilayers using cold atoms in state-dependent optical lattices. Our proposed setup can be used as an alternative platform to explore twisted bilayers which allows one to control the inter/intra-layer coupling in a more flexible way than in the solid-state realizations. We focus on square geometries but also point how it can be extended to simulate other lattices which show Dirac-like physics. This setup opens a path to observe similar physics, e.g., band narrowing, with larger twist angles, to rule out some of the mechanisms to explain the observed strongly correlated effects, as well as to study other phenomena difficult to realize with crystals. As an example of the latter we explore the quantum optical consequences of letting emitters interact with twisted bilayer reservoirs, and predict the appearance of unconventional radiation patterns and emitter interactions following the emergent Moir\'e geometry.	
\end{abstract}

\maketitle

\section{Introduction~\label{sec:intro}}
	
The recent observation of unconventional superconducting~\cite{cao18a} and correlated insulating behaviour~\cite{cao18b} in twisted bilayer graphene has brought the study of twisted van-der Waal materials, that is, rotated weakly coupled layers, at the forefront of condensed-matter research. One of the initial motivations of the twisting was to shift graphene van-Hove singularities closer to half-filling to induce superconductivity by doping~\cite{nandkishore12a,li10a,mcchesney10a}. However, the interest in those systems burst with the observation that the Fermi velocity, $v_F$, around the Dirac point renormalizes strongly for small rotations~\cite{dossantos07a,shallcross10a,trambly10a,morrel10a,bistritzer11a}, especially at the so-called magic angles~\cite{trambly10a,morrel10a,bistritzer11a}. At these angles, the interplay between the interlayer hopping, $\hbar J_\perp$ and the momentum mismatch energy, $\hbar v_F k_\theta$, leads to bands whose bandwidth is comparable to the effective Coulomb interactions and with large density of states, making them an ideal candidate to observe strongly correlated phenomena. This potential has been evidenced by the renowned transport experiments~\cite{cao18a,cao18b} as well as the observation of other exotic effects such as super-planckian dissipation~\cite{cao19a} or ferromagnetism~\cite{sharpe19a}. Beyond these observations, the experimental efforts have been devoted to finding ways of tuning these systems, e.g., using pressure~\cite{yankowitz19a}, electrical gating~\cite{chen18a}, dynamical angle control~\cite{ribeiro18a}, magnetic fields~\cite{kim17a,sharpe19a}, or temperature~\cite{cao19a}, as well as alternative forms to probe the system beyond electron transport~\cite{kerelsky18a,li10a,kim17a,choi19a,yoo18a,ni15a,sunku18a}. 

Since the inter/intra-layer hopping ratio is small in these systems, the magic angles occur for very small rotations, $\theta\lesssim 1^\circ$. This has several consequences: first, it makes the structures more sensitive to structural relaxation~\cite{yoo18a,yankowitz19a}, to disorder, or the effect of the substrate~\cite{sharpe19a}; second, the Moir\'e primitive cells are enormous ($\sim 10^4$ atoms per unit cell), making ab-initio calculations challenging. This is why one of the main theoretical research directions consists in finding accurate descriptions for inter/intra-layer hoppings in tight-binding~\cite{shallcross10a,trambly10a,morrel10a,bistritzer11a,moon12a,fang15a,fang16a} or continuum descriptions~\cite{dossantos07a,sanjose12a}, including lattice relaxation effects~\cite{uchida14a,nam17a,zhang18a,wen18a,carr17a,carr18a,carr18b,koshino18a,guinea19a,walet19a}, as well as searching for minimal models that capture the physics of the so-called active bands~\cite{kang18a,po18b,zou18a,else18a,song18a,yuan18a,delapena17a,classen19a}. Despite all these efforts, the origin of the emergent superconductivity~\cite{lian18,wu18c,peltonen18a,isobe18a,you18a,lin19a,gonzalez18a,po18a,venderbos18a,xu18a,dodaro18a,kennes18a,guinea18a,xu18b,kang18b}, the correlated insulating behaviour~\cite{po18a,pizarro18a,padhi18a,padhi18b}, or even the magic angles~\cite{tarponolsky19a}, is still under debate. For example, several mechanisms have been proposed to explain the experimentally observed superconductivity such as electron-phonon interaction~\cite{lian18,wu18c,peltonen18a}, assisted correlated hopping~\cite{guinea18a,xu18b}, or the effective electron attraction induced by the presence of van-Hove singularities and nested Fermi-surfaces~\cite{isobe18a,you18a,lin19a,gonzalez18a}. Thus, it seems very timely to find other platforms where to study these effects and enlarge our understanding of this exciting system.

In this work, we propose a way to engineer state-dependent optical potentials for cold atoms such that their dynamics mimic that of twisted bilayers. Cold atoms in optical lattices stand nowadays as one of the most mature implementations to simulate condensed-matter problems~\cite{lewenstein07a,bloch12a}, including the physics of graphene-like materials~\cite{tarruell12a} and even their modification when placed on a substrate~\cite{grass16a}. The key ingredient of our proposal is the creation of two orthogonally-polarized standing waves with an angle between them such that when judiciously coupled to an atom they generate twisted optical potentials which trap independently two long-lived atomic states. The localized atomic excitations in each potential can hop within the lattice through the hybridization of their atomic wavefunctions simulating the intralayer hoppings. On top of that, we add an additional global field to connect the two internal atomic states, e.g., via a two-photon Raman transition through an auxiliary level, to obtain the interlayer hopping. Thus, the inter/intra-layer couplings can be controlled independently by optical means. This control opens up the possibility of observing similar phenomena than in solid-state implementations, such as the emergence of narrow bands or the movement (or appearance) of van-Hove singularities, but for larger rotation angles. In this manuscript we focus on square Moir\'e geometries, but it can potentially be extended to brick-wall~\cite{tarruell12a} or hexagonal lattices~\cite{soltan11a}, which would allow the exploration of Dirac-like physics. Finally, we also show how our proposed setup can be adapted to explore physics beyond the one observed with crystals such as the one of emitters interacting with twisted bilayer reservoirs. For that configuration, we predict the appearance of strongly non-Markovian dynamics controlled by the twisting angle, and the generation of emission patterns and interactions following the emergent Moir\'e geometry.

\begin{figure*}[tb]
	\centering
	\includegraphics[width=0.95\textwidth]{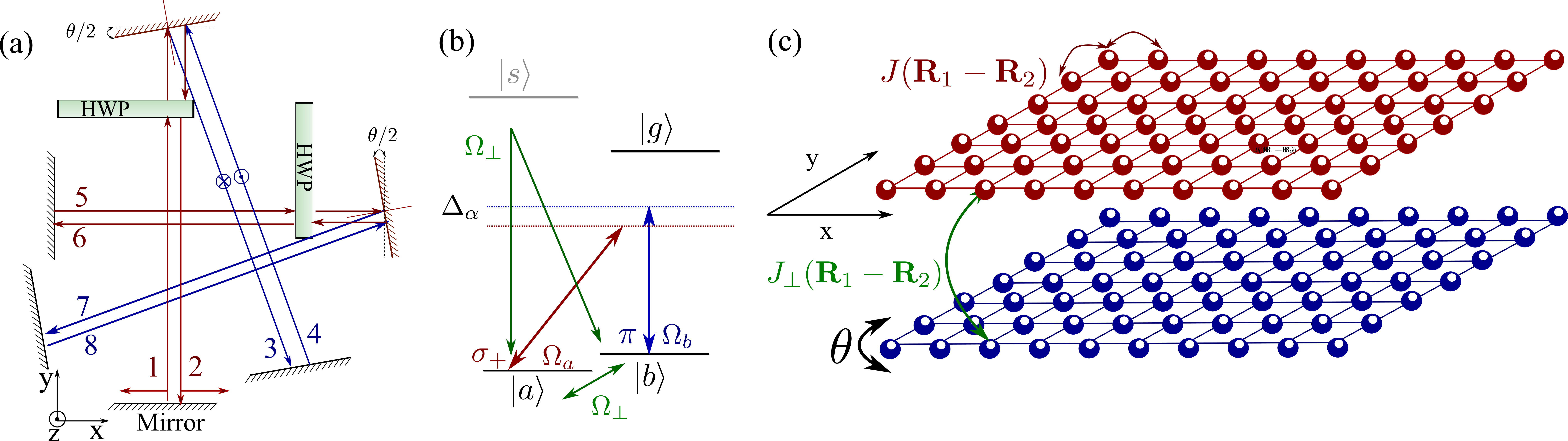}
	\caption{(a) Two movable mirrors plus half-wave plates are used to generate two sets of standing waves with orthogonal polarizations rotated an angle $\theta$ with respect to each other. (b) Possible level structure to harness the orthogonally polarized standing waves to generate an independent optical potential for two long-lived states ($a,b$): we need that these states couple to an additional one $g$ off resonantly, each with an independent polarization of light. In green we depict two possible ways of coupling the $a,b$ states either through a two-photon Raman process via an intermediate state ($s$) or through a direct microwave (optical) transition. (c) Sketch of simulated bilayer trapping of the atoms in $a$-$b$ states depicted in red-blue, respectively. The intra-layer hopping, $J(\mathbf{R}_{12})$, will be controlled through the depth of the optical potentials $V_{a,b}(\RR)$, whereas the inter-layer hopping, $J_\perp(\mathbf{R}_{12})$, can be controlled independently through a direct or indirect transition between the $a,b$ states ($\Omega_\perp$).}
	\label{fig:1}
\end{figure*}

The manuscript is structured as follows: first, in Section~\ref{sec:twist}, we discuss how to generate twisted optical potentials for cold atoms using Alkali-Earth atoms~\cite{ludlow15a}. Then, in Section~\ref{sec:band}, we derive the effective atom dynamics in these twisted optical lattices (in~\ref{subsec:wannier}), and calculate its associated band structure and density of states (in~\ref{subsec:band}). Afterwards, in Section~\ref{sec:qo}, we explain how to simulate quantum optical effects in these setups by using an additional atomic state, and calculate the spontaneous emission dynamics and radiation patterns of an emitter coupled to such twisted bilayer reservoirs. Finally, we summarize our findings in Section~\ref{sec:conclusions}, and provide an outlook of future directions of our work.

\section{Twisted optical potentials~\label{sec:twist}}

In this section we describe in several steps how to obtain twisted independent optical potentials for cold atoms to simulate the physics of twisted crystal bilayers. First, in~\ref{subsec:standing} we explain how to generate a set of orthogonally polarized standing waves with a square geometry and twisted with respect to each other. Second, in~\ref{subsec:ideal} we describe the basic atomic configuration required to harness these standing waves to generate independent optical potentials for two atomic states. Finally, in~\ref{subsec:alkali} we discuss several specific realizations of that configuration based on Alkali-Earth atomic level structure~\cite{ludlow15a}.

\subsection{Twisted orthogonally polarized standing waves~\label{subsec:standing}}
	
The laser configuration is sketched in Fig.~\ref{fig:1}(a): one initially starts with two retro-reflected lasers with in-plane polarization in orthogonal directions ($x$ and $y$) which create a two-dimensional square-like potential $\propto \sin^2(k x)+\sin^2(k y)$ (in red). To create additional standing waves tilted an angle $\theta$ with respect to each of the beams, one can tilt one of the mirrors by an angle $\theta/2$, add a half-wave plate that changes the in-plane polarization $\hat{\epsilon}_x$ into $\hat{\epsilon}_z$ (and viceversa), and include an additional mirror that reflects the light back to create a standing wave. Denoting the light-paths as $1$-$4$ like in Fig.~\ref{fig:1}(a), the momenta and polarization of the lasers along this closed path read $\kk_{1,2}=k(0,\pm 1,0)$, $\kk_{3,4}=k(\pm\sin(\theta),\mp \cos(\theta),0)$, $\hat{\varepsilon}_{1,2}=\pm \hat{e}_x$, and $\hat{\varepsilon}_{3,4}=\pm \hat{e}_z$, where $\lambda=2\pi/k$ is the wavelength of the laser. Since the polarizations in the paths $1$-$2$ and $3$-$4$ are orthogonal, the two standing waves will not interfere even if they have the same frequency. In the orthogonal path, denoted by $5$-$8$ in the Fig.~\ref{fig:1}(a), we have something similar with $\kk_{5,6}=k'(\pm 1,0,0)$, $\kk_{7/8}=k'(\mp\cos(\theta),\mp \sin(\theta),0)$, $\hat{\varepsilon}_{5,6}=\pm \hat{e}_y$, and $\hat{\varepsilon}_{7,8}=\pm \hat{e}_z$. To avoid interference between the paths $3,4,7,8$ (and $1,2,5,6$) that have the same polarization, we use the standard procedure~\cite{grimm00a} of sending slightly detuned lasers with  $\lambda\approx \lambda'$ (so that the periodicities of the potentials are the same) but whose frequency $\omega_L(')=2\pi/\lambda(')$ are sufficiently far such that their cross-talk average out in the atomic timescales (of the order tens of kHz). Taking into account all these considerations, the time averaged intensity felt by a linear/circularly polarized atomic transition read:
\begin{align}
\label{eq:Esigma}
|\hat{e}_{\sigma_{\pm}}\cdot \EE(\RR)|^2&=\frac{|E_0|^2}{2} I_{\sigma_{\pm}}(\RR)\,,\\
|\hat{e}_{z}\cdot \EE(\RR)|^2&=|E_0|^2 I_{\pi}(\RR;\theta)\,,\label{eq:Ez}
\end{align}
where $|E_0|^2$ is the intensity of the lasers generating the standing waves, and the functions $I_{\alpha}$ are the (adimensional) electric field profile intensity that read:
\begin{align}
I_{\sigma_\pm}(\RR)=&\sin^2(k x)+\sin^2(k y)\,,\\
I_z(\RR;\theta)=&\sin^2(k(x \sin(\theta)-y\cos(\theta)))\nonumber \\
+&\sin^2(k(x \cos(\theta)+y\sin(\theta)))\,,
\end{align}
where we have defined $z$ as the quantization axis, and written the in-plane electric field as a combination of circular polarized light: $\hat{e}_{x}=\frac{1}{\sqrt{2}}\left(\hat{e}_{\sigma_{-}}+\hat{e}_{\sigma_{+}}\right)$ and $\hat{e}_y=\frac{i}{\sqrt{2}}\left(\hat{e}_{\sigma_{-}}-\hat{e}_{\sigma_{+}}\right)$. Notice the factor $1/2$ which appears in the circular component intensity in Eq.~\ref{eq:Esigma} because the in-plane intensity $\hat{e}_{x/y}$ divides between the two circular polarizations. The vertical confinement can be obtained adding another pair of retro-reflected lasers in the vertical direction, $k_{9/10}=k_z(0,0,\pm 1)$, that will generate a standing wave $\propto \sin^2(k_z z)$. Since we do not want the atoms to hop in that direction we can choose a different frequency from the one generating the twisted standing waves such that no matter the polarization of $\hat{\varepsilon}_{9/10}$ it does not interfere with them.

\subsection{Basic atomic scheme~\label{subsec:ideal}}

Let us now explain how by adequately choosing the atomic level structure, this set of standing waves forms a different optical potential for two different long-lived atomic states labeled as $a,b$. We consider here the configuration depicted in Fig.~\ref{fig:1}(b) where these two states connect to an additional one $g$ (with decay rate $\Gamma_g$) through an orthogonal polarization of light, e.g., $\sigma_{+},\pi$ (we show in the next section how). Consequently, the effective Rabi amplitude of these optical transitions, $\Omega_{a,b}(\RR)$, inherit the spatial dependence of the intensity profiles of Eqs.~\ref{eq:Esigma}-\ref{eq:Ez}. This can be explicitly shown by writing $\Omega_{a,b}(\RR)=\tilde{\Omega}_{a,b} I_{\sigma_{+},\pi}(\RR)$, where in $\tilde{\Omega}_{\alpha}$ we include the overall transition strength  including the contribution of the Clebsch-Gordan coefficients as well as the $1/2$ factor of Eq.~\ref{eq:Esigma}. Next, we assume that these long-lived states are split in energy by $\delta=\omega_b-\omega_a$, e.g., by a magnetic field, such that their optical transitions to $g$ have different detunings $\Delta_{\alpha}=\omega_g-\omega_L-\omega_\alpha$, for $\alpha=a,b$.

 Then, in the limit where $|\tilde{\Omega}_{\alpha}|\ll \Delta_{\alpha}$ and $|\tilde{\Omega}_{a}\tilde{\Omega}_{b}|\ll \Delta_{\alpha}\delta$, the dressing of the states $\alpha$ with $g$ leads to the following state-dependent optical potential~\cite{grimm00a}:
\begin{align}
V_{\alpha}(\RR)\approx-\frac{|\tilde{\Omega}_{\alpha}|^2}{\Delta_\alpha}I_\alpha(\RR)\,.
\end{align}

The role of the energy splitting $\delta$ is two-fold:
\begin{itemize}
	\item It allows to compensate for the different coupling amplitudes of the optical transitions generating the potentials. For example, if we want to obtain the same potential depth $V_D$ for the $a/b$ levels we must impose:
	\begin{align}
	\label{eq:cond1}
	\frac{|\Omega_a|^2}{\Delta_a}=\frac{|\Omega_b|^2}{\Delta_b}\equiv V_D\,.
	\end{align}
		
	As we will see afterwards, this guarantees obtaining the same hopping rate in each optical potential, which is what simulates the intralayer hopping $J$ for both atomic excitations.
	
	\item It suppresses the unwanted two-photon transitions between the states. The two-photon amplitude between two states scales with $ \Omega_{\mathrm{2ph}} \approx \tilde{\Omega}_a\tilde{\Omega}_{b}^*/\Delta_a$, which is of the order of the potential depth ($V_D$), whereas its effective detuning is $\delta$. Thus, for the two-photon transition probability ($\varepsilon_\mathrm{2ph}$) to be small we require:
	\begin{align}
	\label{eq:cond2}
	\varepsilon_\mathrm{2ph}\sim \left(\frac{V_D}{\delta}\right)^2\ll 1\,,
	\end{align}
	
\end{itemize}

Apart from these trapping lasers, to complete the proposal we need a mechanism that couples the $a,b$ states directly. Depending on the particular choice for the $a,b$ levels (see next section) this can be done in several ways as depicted in green in Fig.~\ref{fig:1}(b), either with a global microwave (optical) field which couples directly these states, or with a two-photon Raman process through an auxiliary excited state. Any of these mechanisms provides an effective coupling strength, $\Omega_\perp$, between the $a/b$ states that can be controlled independently from the lasers generating the twisted optical potentials. Thus, we have a knob to change the interlayer coupling between the states, $J_\perp$, which is independent from the one generating the intralayer hopping, $J$, as schematically depicted in Fig.~\ref{fig:1}(c).

Before giving a particular realization of this atomic level structure, let us discuss more conditions values for the proposal:
\begin{itemize}

	\item The trapping depths must be big enough such that the atoms are trapped within their motional ground state. The rule-of-the thumb condition is that it should be bigger than the recoil energies~\cite{bloch08a} of the lattice $V_D>E_R=\frac{\hbar^2 k^2}{2 m}$, of the order of $1-10$ kHz for standard cold atoms experiments. Thus, $V_D$ should be $\sim 10$-$100$ kHZ.
	
	\item If $g$ is an unstable state with a decay rate $\Gamma_g$, the $a,b$ levels acquire a finite decoherence rate which scales as:
	\begin{align}
	\Gamma_*\approx \frac{V_D}{\Delta}\Gamma_g\,, 
	\end{align}
	and that will set a limitation on the coherence timescales of our simulation. In practice, $\Gamma_*$ should be much smaller that the effective atom dynamics that we want to simulate, e.g., the band-width of the emergent Moir\'e bands. This suggests using states for which $\Gamma_g$ is as small as possible, as we discuss in the next section.
\end{itemize}

\subsection{Specific realizations using Alkali-Earth atoms~\label{subsec:alkali}}

The ideal level structure considered in Fig.~\ref{fig:1}(b) can be easily engineered in an optical transition between a ground state with total electron angular momentum $J=1$ and an excited state with $J'=0$. These optical transitions appear in rare-earth atoms like Samarium~\cite{nottelmann93a}. Unfortunately, these atoms are difficult to cool down to their ground state. Other ways of generating such state-dependent optical traps consists in using ground and excited hyperfine levels in Alkali atoms~\cite{rubio18a,krinner18a} for the $a/b$ and $g$ states, respectively. However, these potentials typically suffer from photon loss decoherence due to the small hyperfine energy splitting so that the conditions required above would be hardly fulfilled.

For the above reasons we just provide several possible realizations of the scheme of Fig.~\ref{fig:1}(b) using Alkali-Earth atoms~\cite{ludlow15a}, such as Calcium (Ca), Strontium (Sr), or Ytterbium (Yb). All these atoms share an electronic configuration with two electrons in an outer $s$-shell, generating a level structure (see Fig.~\ref{fig:AlkaliEarth}) with several remarkable features:

\begin{figure}[tb]
	\centering
	\includegraphics[width=0.4\textwidth]{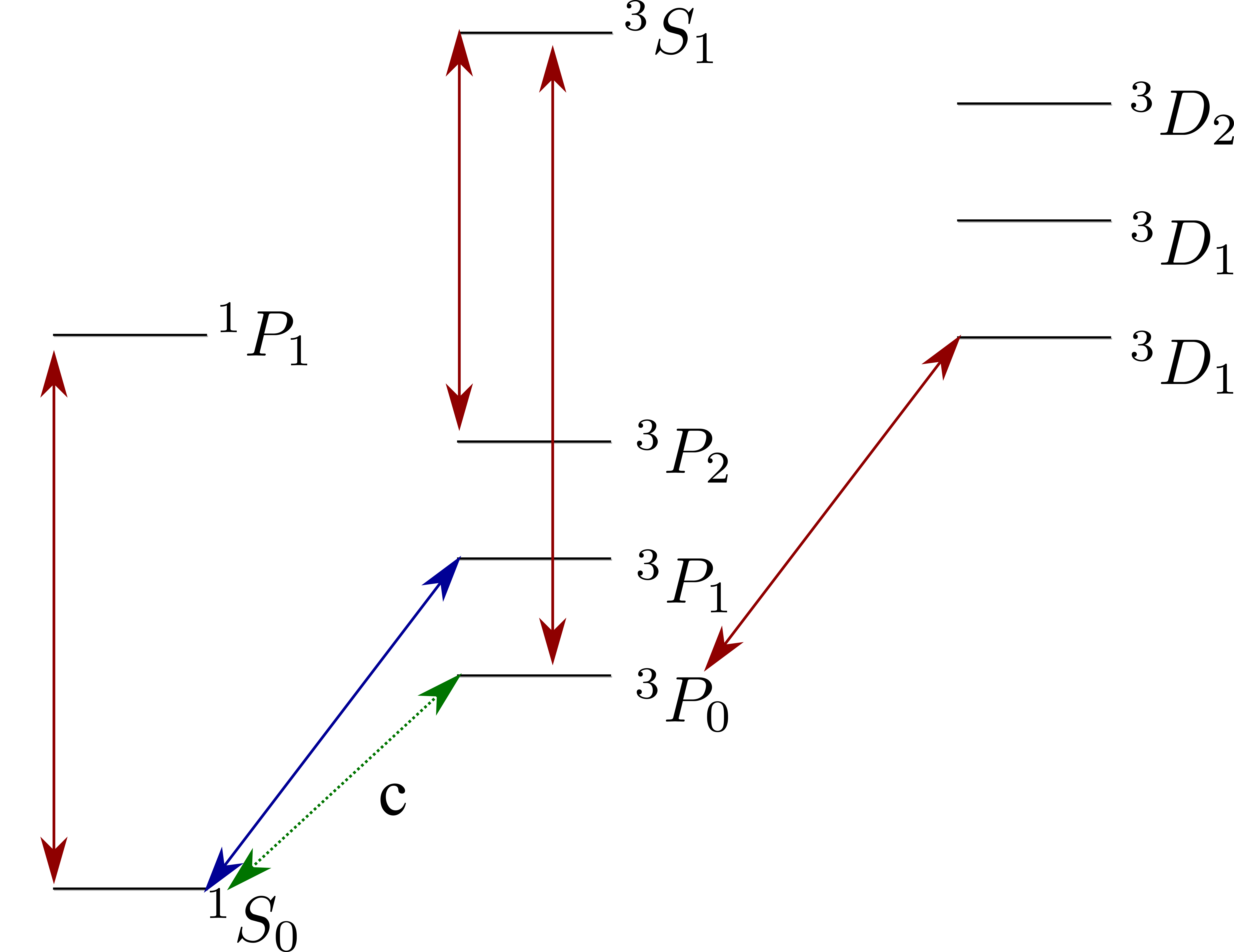}
	\caption{Electronic structure of Alkaline-Earth atoms with the most relevant optical transitions highlighted: in red the strong singlet-triplet transition and in blue/green the simply (doubly) spin forbidden transition. We use the standard spectroscopic notation for the levels $^{2S+1} L_J$, where $S$ is the total electron spin number, $L$ denotes the electronic angular momentum, and $J=L+S$ is the total electron angular momentum. }
	\label{fig:AlkaliEarth}
\end{figure}
\begin{itemize}
	\item The two electrons can align in singlets ($S=0$) and triplets ($S=1$). Strong transitions (depicted in red in Fig.~\ref{fig:AlkaliEarth}) occur between singlet-triplet states, e.g., $^1 P_1$-$^1 S_0$ with decay rates $\Gamma/(2\pi)\sim 30$ MHz for Sr or Yb~\cite{ludlow15a}. These transitions are well separated spectrally and can be used for trapping, state detection, or cooling. In fact, for all these atoms there exist already refined laser cooling techniques which can bring them to their ground state~\cite{ludlow15a,takasu03a,kraft09a,stellmer09a,snigirev17a,riegger18a}.
	
	\item Weaker decay rates occur for spin-forbidden transitions. For example, the transition  $^3 P_1$-$^1 S_0$ (in blue) is a simply spin forbidden transition leading to lifetimes of the order of $300$ Hz for Ca~\cite{kraft09a}, $7.5$ kHz for Sr~\cite{ludlow15a}, and $180$ kHz for Yb~\cite{ludlow15a}. The clock transition $^3 P_0$-$^1 S_0$ is doubly forbidden (in green) leading to narrow linewidths of $\sim 1$-$10$ mHz for Ytterbium and Strontium, respectively~\cite{ludlow15a}, whereas the state $^3 P_2$ has even longer predicted lifetime. 
	
	\item All these atoms have bosonic ($^{40}$Ca, $^{88}$Sr, $^{174}$Yb) and fermionic ($^{43}$Ca, $^{87}$Sr, $^{171,173}$Yb) stable isotopes. The fermionic ones have non-zero nuclear spin $I\neq 0$ depending on their atomic mass, e.g., $I=1/2$ and $5/2$ for $^{171}$Yb and $^{173}$Yb, respectively. This nuclear spin can play an important role in the context of quantum simulation and information because it leads to degenerate multiplet hyperfine levels in the clock states~\cite{daley08a,gorshkov10a}. Moreover, using a magnetic field these levels can be split with differential Zeeman shifts for the $^1 S_0$ and $^3 P_0$ of the order of $1$ MHz/T~\cite{ludlow15a}, which allows to address them independently~\cite{daley08a}.
\end{itemize}

\begin{figure*}[tb]
	\centering
	\includegraphics[width=0.94\textwidth]{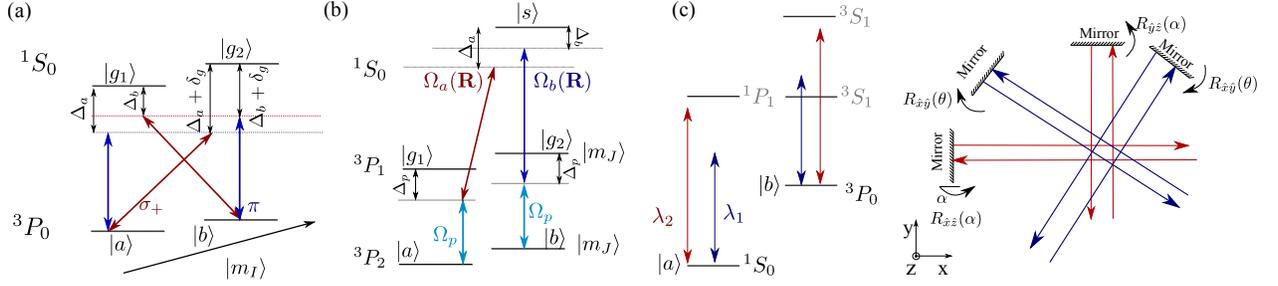}
	\caption{ (a) Relevant energy levels and optical transitions for the proposal using the hyperfine levels of the clock state $^3P_0$ discussed in section~\ref{subsub:hyper}. To compare with the ideal scheme of Fig.~\ref{fig:1}(b) we plot the hyperfine levels $g_{1,2}$ of $^1 S_0$ above the ones of $^3 P_0$ even though they have lower energy.  (b) Relevant energy levels and optical transitions for the proposal using the fine structure levels of $^3 P_2$ discussed in section~\ref{subsub:fine}. Like in panel (a), we plot the level structure inverted in energy to help to compare to the ideal scheme of Fig.~\ref{fig:1}(b). (c) Relevant energy levels, optical transitions, and mirror configuration for the implementation where the clock states $^1S_0, ^3 P_0$ are the $a,b$ states. We use the turnout wavelengths discussed in section~\ref{subsub:turnout} to generate two independent optical potentials. One of the lasers must be sent out of plane such that the in-plane momenta is the same for both fields. The $R_{ij}(\beta)$ in the figure denotes the rotation of the mirror in the $ij$-plane with an angle $\beta$.}
	\label{fig:schemes}
\end{figure*}

After reviewing the main spectral properties of these atoms, let us propose three independent ways of generating the twisted optical potentials as summarized in Fig.~\ref{fig:schemes}: the first two (in panels a-b) are inspired in the scheme discussed in the previous Section and relies on a judicious Zeeman splitting of hyperfine or fine structure levels using magnetic fields. On the other hand, the third one (panel c) does not rely on polarization or magnetic fields but on sending one of the lasers with different frequency and a certain out-of-plane angle with respect to the plane of the layer, as proposed in the context of quantum information in Ref.~\cite{daley08a}.

\subsubsection{Using hyperfine levels of the clock state $^3 P_0$~\label{subsub:hyper}}

One possible realization consists in using the hyperfine levels of the state $^3 P_0$ of a fermionic isotope as our $a/b$ levels. In particular, we can take $\ket{a}=\ket{^3 P_0,I,m_I=-I}$ and $\ket{^3 P_0,I,m_I=-I+1}$. Differently from our ideal scheme, the off-resonant lasers $\Omega_{a/b}$ connect now these states to two different hyperfine levels of the ground state, namely, $\ket{g_1}=\ket{^1 S_0,I,m_I=-I}$ and $\ket{g_2}=\ket{^1 S_0,I,m_I=-I+1}$ (see Fig.~\ref{fig:schemes}(a)). Thus, both the $a,b$ states feel the linear and circularly polarized intensities, but with different detunings such that they yield the following optical potentials:
\begin{align}
\label{eq:nondes}
V_{a}(\RR)\approx- \frac{|\Omega_a|^2}{\Delta_{a}} I_\pi(\RR)-\frac{|\Omega_b|^2}{\Delta_{a}+\delta_g} I_{\sigma_+}(\RR)\,,\\
V_{b}(\RR)\approx - \frac{|\Omega_b|^2}{\Delta_{b}} I_{\sigma_-}(\RR)-\frac{|\Omega_a|^2}{\Delta_{b}+\delta_g} I_\pi(\RR)\,, \label{eq:nondes2}
\end{align}
where we have defined $\Delta_{a/b}=\omega_{g_1}-\omega_{a/b}-\omega_L$, and $\delta_g=\omega_{g_2}-\omega_{g_1}$~\footnote{Notice that here for simplicity we assume that the Clebsch Gordan coefficients for the transitions $a\rightarrow g_2$ and $b \rightarrow g_1$ (as well as for $a\rightarrow g_1$ and $b\rightarrow a_1$) are the same. Even if they are different this will become irrelevant in the regime where one can neglect the non-ideal contributions.}.  Imposing~Eq.~\ref{eq:cond1} such that the potential depth $V_D$ for both potentials is the same, we can rewrite the potentials:
	\begin{align}
	\label{eq:nondes3}
	V_{a}(\RR)\approx -V_D\left( I_\pi(\RR)+\frac{\Delta_{b}}{\Delta_{a}+\delta_g} I_{\sigma_+}(\RR)\right)\,,\\
	V_{b}(\RR)\approx -V_D\left(I_{\sigma_-}(\RR)+\frac{\Delta_{a}}{\Delta_{b}+\delta_g} I_\pi(\RR)\right)\,,	\label{eq:nondes4}
	\end{align}

Notice, that $V_{a,b}(\RR)$ contains now two contributions: the desired one (left hand side of the bracket) plus a correction (right-hand side) that might spoil the twisted bilayer behaviour. Thus, in order for the right-hand term of these equations to be negligible, we require that $\frac{\Delta_{\alpha}}{\Delta_{\beta}+\delta_g}\ll 1$. Since $\delta_g\sim 2$ MHz for $B\sim 1$ T, this restricts $\Delta_{\alpha}/(2\pi)\sim 0.2$ MHz. The advantage here is that since the dressing is done with an stable state ($^1 S_0$), ones does not have to worry about canceling $\Gamma_g$. Thus, $\Omega_{\alpha}$ can be a significant fraction of $\Delta_\alpha$, e.g., $\Omega_\alpha\approx 0.25\Delta_\alpha$, yielding $V_D/(2\pi)\sim 10$ kHz. 

\subsubsection{Using fine structure levels of the $^3P_2$ through a two-photon transition\label{subsub:fine}}

One of the challenges of the previous realization is that the use of hyperfine states limits the energy splitting $\delta_g$ required to cancel the non-desirable terms in Eqs.~\ref{eq:nondes3}-\ref{eq:nondes4}. An alternative is to use fine structure levels as our target levels, e.g., $\ket{a,b}=\ket{^3 P_2, m_J=-1,0}$, whose Zeeman splitting can be several orders of magnitude larger for similar magnetic fields ($\sim10$ GHz for $\sim T$ fields). The problem here is that these states do not connect directly to the ground state $^1 S_0$. Thus, we have to do it through a two-photon transition via the states $\ket{g_{1,2}}=\ket{^3 P_1, m_J=-1,0}$ as depicted in Fig.~\ref{fig:schemes}(b): the twisted standing wave lasers of section~\ref{subsec:standing} couple the ground state $^1 S_0$ to the intermediate $g_{1,2}$ states through the spatially dependent $\Omega_{a,b}(\RR)$ lasers. This generates a state-dependent Stark-shift on the $g_{1,2}$ levels, which translates to the $^3 P_2$ through a global laser field with linear polarization, Rabi amplitude $\Omega_p$, and frequency $\omega_{L,p}$. This laser couples the $^3 P_1$ and $^3 P_2$ states off resonantly with an effective detuning which in this case is approximately equal for both transitions $\Delta_p=\omega_{g_1}-\omega_{a}-\omega_{L,p}\approx \omega_{g_2}-\omega_{b}-\omega_{L,p}$~\footnote{This is a good approximation because the $^3 P_J$ states have $S=L=1$ such that the Land\'e factor $g_J\approx 3/2$ independent on the total angular momentum}, whereas the global two-photon transition detuning for each path reads $\Delta_{a,b}=\omega_{s}-\omega_L-\omega_{L.p}-\omega_{a,b}$. This two photon transition can be shown to yield a spatially dependent potential for the $a,b$ states which reads:
\begin{align}
\label{eq:stateStark4}
V_{a,b}(\RR)\propto -\frac{|\Omega_p|^2}{\Delta_p^2}\frac{|\tilde{\Omega}_{a,b}|^2}{\Delta_{a,b}} I_{\sigma_{-},\pi}(\RR)\,,
\end{align}
when $|\tilde{\Omega}_{a,b}|^2\ll \Delta_{a,b}\Delta_p$. Notice that we require that the laser $\Omega_p$ does not introduce any additional spatial dependence which can be achieved with a running-wave laser configuration. Thus, fixing $\Delta_{a,b}$ like in Eq.~\ref{eq:cond1}, the overall trapping depth scales in this case as:
\begin{align}
V_{D}=\frac{|\Omega_p|^2}{\Delta_p^2}\frac{|\tilde{\Omega}_{a,b}|^2}{\Delta_{a,b}}\,.
\end{align}

Since the intermediate states $g_{1,2}$ have a finite lifetime $\Gamma_g^{-1}$, the $a/b$ states acquire a decay rate which scales with $\Gamma_*\sim \frac{|\Omega_p|^2}{\Delta_p^2}\Gamma_g$. Thus, to obtain coherence times of the order of seconds, we must impose $|\Omega_p/\Delta_p|\sim 10^{-1},10^{-2}$ for Ca or Sr atoms whose $\Gamma_{g}/(2\pi)=0.3,7.5$ kHz, respectively. Since $\Delta_{a,b}$ can be $\sim 10$ GHz in this case, and $|\tilde{\Omega}_{a,b}|\ll \Delta_{a,b}$, this leads to $V_D/(2\pi)\sim 10$-$1000$ KHz, depending on the ratio $\Omega_p/\Delta_p$ required to cancel $\Gamma_g$. 

The main technical difficulty of this proposal is to drive directly the transition $^3 P_1$-$^3 P_2$ which requires a laser in the $10$ THz range. An alternative way might be driving indirectly with a two-photon Raman transition with optical lasers through the $^3 S_1$ state. In that case, one must use a detuning large enough to cancel the decay rate of the $^3 S_1$ state.

\subsubsection{Turnout wavelengths with angled beams~\label{subsub:turnout}}

Finally, let us mention one last possibility which does not rely on the generation of the orthogonally polarized standing waves explained in section~\ref{subsec:standing}. It is inspired on the existence of the so-called turnout wavelengths ($\lambda_{1,2}$) of Alkaline-Earth atoms~\cite{daley08a} in which the polarizabilities of the two-long-lived states $\ket{^1 S_0}=\ket{a}$ and $\ket{^3 P_0}=\ket{b}$ vanish independently, that is, $\alpha_{a,b}(\lambda_{1,2})=0$ and $\alpha_{a,b}(\lambda_{2,1})\neq 0$. This allows one to create two independent potentials for the $a/b$ states using two lasers with these wavelengths. The only problem for our purposes is that these wavelengths differ significantly, e.g., $\lambda_{1,2}=689,627$ nm for Strontium~\cite{ludlow15a}, such that if both lasers are sent in-plane as in Fig.~\ref{fig:1}(a) their potentials will have different periodicities. However, this can be compensated by sending the shorter wavelength laser with an out-of-plane angle $\alpha$~\cite{daley08a}, as schematically depicted in Fig.~\ref{fig:schemes}(c), such that the in-plane momenta is reduced by a factor $\cos(\alpha)=\lambda_2/\lambda_1$.  For the vertical trapping, one can send an additional pair of retro-reflected lasers in the $z$ direction with the so-called magic-wavelength ($\lambda_m$) that leads to the same polarizability for both states  $\alpha_{^1 S_0}(\lambda_{m})=\alpha_{^3 P_0}(\lambda_{m})$~\cite{ludlow15a}, and with a large intensity such that it provides a much deeper potential than the one in the XY plane. The coupling between the $a,b$ levels can be obtained by resonantly driving the clock transition or through an intermediate state, e.g., $^3 S_1$, to avoid the small dipole moment of the clock transition.

Let us finally provide some estimates of the coherence times and trapping depths of this configuration. Since the metastable states $a,b$ are dressed by the excited states $^1 P_1$ and $^3 S_1$, respectively, with lifetimes $\Gamma_e/(2\pi)\sim 10$-$100$ MHz, for the coherence times of $\Gamma_{a/b}/(2\pi)\lesssim 1$ Hz the dressing with the excited states have to be kept small~$\tilde{|\Omega}_{a,b}/\Delta_{a,b}|\sim 10^{-3}$-$10^{-4}$. Thus, for $\Delta_{a,b}$'s of the order of $100$ GHz, one can obtain $V_D/(2\pi)\sim 1$-$100$ kHz.

\section{Emergent band structure and density of states~\label{sec:band}}

Here, in Section~\ref{subsec:wannier} we write the Hamiltonian describing the dynamics of the atoms moving in the twisted optical potentials $V_{a/b}(\RR)$. Then, in Section~\ref{subsec:band} we calculate the associated density of states of the emergent band structure. For completeness, in Section~\ref{subsec:dirac} we discuss the possibility to go to other geometries to obtain Dirac-like physics.

\subsection{Effective hopping dynamics in the twisted optical potentials: intra- and interlayer couplings~\label{subsec:wannier}}

The dynamics of atoms subject to the state-dependent optical potential $V_{a,b}(\RR)$ plus the coupling between the internal atomic states $a,b$ through $\Omega_\perp$ can be described by the following Hamiltonian~\cite{jaksch98a}:
\begin{align}
H_B&=\sum_\alpha\int d\RR\hat{\psi}_\alpha^\dagger(\RR)\left[-\frac{\hbar^2}{2 m}\nabla^2+V_\alpha(\RR)\right]\hat{\psi}_\alpha(\RR) \label{eq:hop}\\
      & + \sum_{\alpha} \frac{4\pi\hbar^2 a_{s,\alpha}}{2m}\sum_\alpha\int d\RR \hat{\psi}^\dagger_\alpha(\RR) \hat{\psi}^\dagger_\alpha(\RR) \hat{\psi}_\alpha(\RR) \hat{\psi}_\alpha(\RR) \label{eq:U}\\
      & + \Omega_\perp \int d\RR e^{i\kk_L\cdot \RR-i\omega_L t}\hat{\psi}_a(\RR)\hat{\psi}^\dagger_b(\RR)+\mathrm{H.c.}\,, \label{eq:Hab}
\end{align}
where $\hat{\psi}_\alpha(\RR)$ is the atom field operator of the state $\alpha$, that will be fermionic or bosonic depending on the isotope we are considering. Let us now explain in detail the different parts of the Hamiltonian:
\begin{itemize}
\item \emph{Hopping dynamics (Eq.~\ref{eq:hop})}. This term corresponds to the kinetic energy of the atoms plus the optical potential in each internal atomic state.  When the trapping depth is big enough ($V_D>E_R$), the atoms localize within the minima of the potential, that we denote by $\RR_{\alpha,\nn}$ with $\nn\in\mathbb{Z}^2$ and $\alpha=a,b$, and which can be described as parabolic isotropic traps~\footnote{To obtain accurate descriptions beyond this limit one needs to calculate the exact band-structure using Bloch theorem and obtain the maximally localized Wannier functions~\cite{marzari97a}. In any case, the Hamiltonian parameters, e.g., tunneling, can also be controlled through the optical potential features.}. Notice that here we are already considering that the trapping in the $z$ direction is much deeper than in the $x$-$y$ plane such that the atoms can only hop in these directions. Within that regime, we can expand the atomic field operators in a separable Wannier basis in the $x$-$y$ direction, and keep only the lowest ground state levels such that $\hat{\psi}_\alpha(\RR)=\sum_{\nn}w_\alpha(\RR-\RR_{\alpha,\nn}) \alpha_\nn$, with
\begin{align}
w_\alpha(\RR-\RR_{\alpha,\nn})=\frac{1}{L_\alpha \sqrt{\pi}} e^{-\frac{x^2+y^2}{2 L_\alpha^2}}=\frac{1}{L_\alpha \sqrt{\pi}} e^{-\frac{r^2}{2 L_0^2}}\,.
\end{align}
with $L_\alpha=\sqrt{\frac{\hbar}{m \omega_{t,\alpha}}}$ being the ground state wavefunction size, $\hbar\omega_{t,\alpha}$ the trapping frequency. The $\alpha_\nn$ ($\alpha^\dagger_\nn$) represent the annihilation (creation) of an atomic excitations at site $\RR_{\alpha,\nn}$. Projecting this part of the Hamiltonian in the Wannier basis we arrive to the following form:
\begin{align}
H_\alpha=-\sum_{\nn,\mm} J_\alpha(\nn-\mm) \alpha_\nn^\dagger \alpha_\mm\,,
\end{align}
 where the tunnelings between the different sites are given by the following overlap integrals~\cite{bloch08a}:
\begin{align}
J_\alpha(\nn-\mm)&=
 \int d\RR w_\alpha^*(\RR-\RR_{\alpha,\nn})\left[-\frac{\hbar^2}{2 m}\nabla^2+V_\alpha(\RR)\right]\times \nonumber
\\ &w_
 \alpha(\RR-\RR_{\alpha,\mm})\propto E_R e^{-|\RR_{\alpha,\nn}-\RR_{\alpha,\mm}|^2/(4 L_\alpha^2)}\,,
\end{align}
while for $\nn=\mm$ is equal to $\omega_{t,\alpha}$. When the trapping depth of the potentials $V_{a,b}(\RR)$ is the same, so are the trapping frequencies ($\omega_{t,\alpha}\equiv \omega_t$)  and ground state sizes ($L_\alpha\equiv L_0$), which ultimately determine the strength and range of the hoppings, respectively. Thus, the hopping within each simulated layer will be the same $J_\alpha(\nn)\equiv J(\nn)$.

\item  \emph{On-site interactions (Eq.~\ref{eq:U})}. In the low energy limit these interactions are well approximated by $s$-wave scattering whose strength is determined by so-called scattering length $a_{s,\alpha}$~\cite{bloch08a}. In the Wannier basis, this part of the Hamiltonian reads:
\begin{align}
H_{U}=\sum_{\alpha,\nn} \frac{U_\alpha}{2} \alpha^\dagger_\nn \alpha_\nn  (\alpha^\dagger_\nn \alpha_\nn-1)\,,
\end{align}
with $U_\alpha=\frac{4\pi\hbar^2 a_{s,\alpha}}{m}\int d\RR |w_\alpha(\RR)|^4$. One of the advantage of our setup is that the scattering length, $a_{s,\alpha}$ (and consequently $U_\alpha$) can be controlled through Feshbach resonances~\cite{bloch08a}. Thus, one can tune from having purely non-interacting models with $U_\alpha\approx 0$ to strongly interacting ones of repulsive or attractive character. Furthermore, with the proper choice of the atomic levels one can also obtain crossed on-site interactions between the different internal atomic states $ \propto U_{ab} a^\dagger_\nn a_\nn b^\dagger_\nn b_\nn$. 

\item \emph{Direct $a,b$ coupling (Eq.~\ref{eq:Hab}).} Here, $\Omega_\perp$ is the overall strength of the field connecting the states $a$ and $b$, $\kk_L$ is the in-plane projection of the momentum of the field mediating the transition, and $\omega_L$ its frequency. For simplicity, we can set $\omega_L\equiv 0$ and choose $\kk_L=(0,0)$ (assuming it is coming perpendicular to the layers). With these simplifications, and expanding in the Wannier basis, the $H_{ab}$ Hamiltonian reads~\cite{devega08a,navarretebenlloch11a}:
\begin{align}
\label{eq:inter}
H_{ab}=  - \sum_{\nn,\mm} \left(J_\perp(\RR_{a,\nn}-\RR_{b,\mm}) a^\dagger_\nn b_\mm+\mathrm{H.c.}\right)\,,
\end{align}
whose spatial dependence will be same than for the intralayer couplings:
\begin{align}
J_\perp(\nn-\mm)\propto \Omega_\perp e^{-|\RR_{a,\nn}-\RR_{b,\mm}|^2/(4 L_0^2)}\,,
\end{align}
since it comes from the overlap of the Wannier states between two sites, but whose strength can be tuned independently through $\Omega_\perp$.

\end{itemize}

Summing up, the global Hamiltonian describing the atom dynamics reads:
\begin{align}
\label{eq:bath}
H_B=\sum_\alpha H_\alpha+H_{U,\alpha}+H_{ab}\,.
\end{align}
where we can control: i) the inter/intra-layer hopping ratios ($J_\perp/J$) with $V_D$ and $\Omega_\perp$ as well as their range (through $L_0$); ii) the strength of the on-site atomic interactions $U_{\alpha}, U_{ab}$ through Feshbach resonances; iii) the statistics of the particles with the atomic isotope. Even though we will not consider it for this manuscript, the range of the atomic interactions can be extended by dressing the $a,b$ states with a Rydberg level~\cite{gallagher05a}.  Thus, we have a very unique platform to explore and simulate the physics of fermionic and bosonic twisted bilayers. 

In the next section, we illustrate the potential of the proposed setup to simulate similar phenomena than in twisted van der Waals materials, in terms of observation of narrow bands or higher-order van-Hove singularities. For that, we calculate the emergent band structure and density of states for the twisting angles in which the two layers have a crystalline structure (Moir\'e patterns). 

\begin{figure}[tb]
	\centering
	\includegraphics[width=0.45\textwidth]{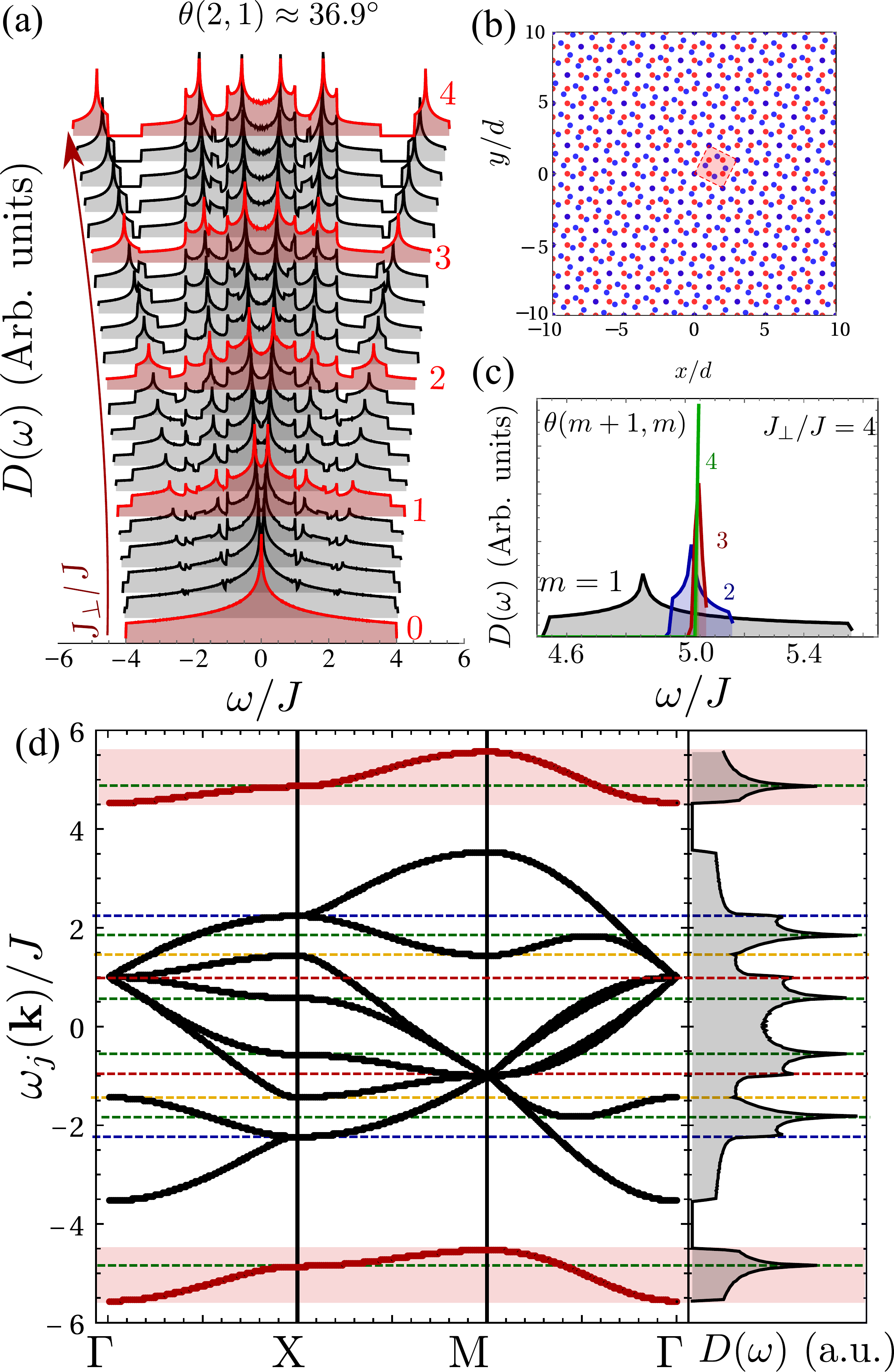}
	\caption{(a) Density of states of a square Moir\'e twisted bilayer with an angle $\theta(2,1)$ as defined in Eq.~\ref{eq:angle} for several equally spaced $J_\perp/J$ from $0$ to $4J$. (b) Associated Moir\'e pattern for $\theta(2,1)$, where we plot in red/blue the atomic sites in the $a/b$ potentials. (c) Zoom of the density of states of the isolated upper band for fixed $J_\perp/J=4$, and for several angles $\theta(m+1,m)$ as depicted in the legend. Larger $m$'s correspond to smaller angles and yield higher density of states and narrower bandwidths. (d) Projected band structure $\omega_j(\kk)/J$ for the path defined along the symmetry points $\Gamma$-X-M-$\Gamma$ together with the associated density of states for $\theta(2,1)$ and $J_\perp/J=4$.}
	\label{fig:2}
\end{figure}
\begin{figure}[tb]
	\centering
	\includegraphics[width=0.4\textwidth]{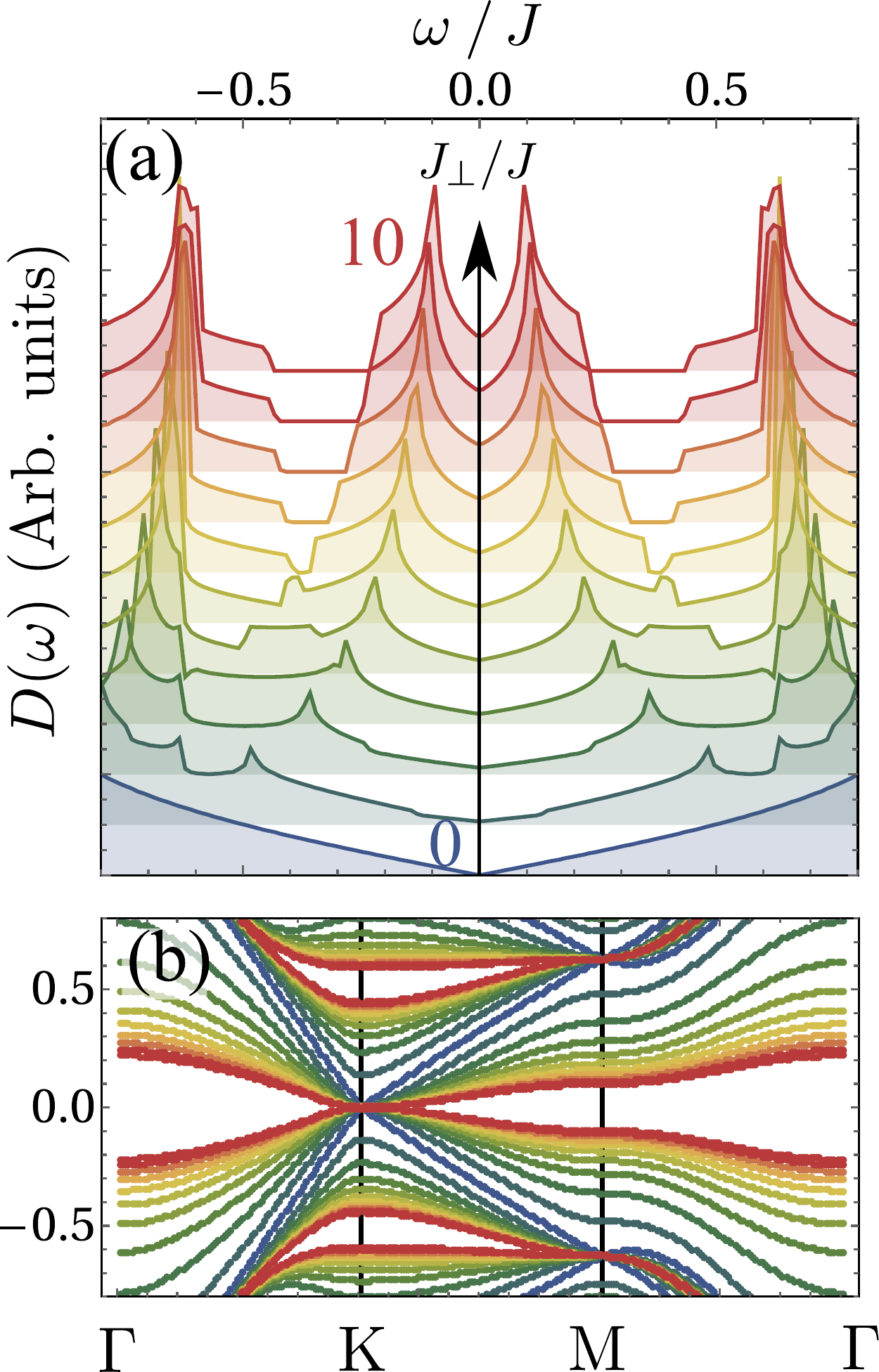}
	\caption{(a) Density of states and band structure (b) for a twisted honeycomb bilayer lattice with an angle $\theta_\mathrm{hc}(2,1)$. We zoom around the Dirac frequency of the individual layers and plot for several $J_\perp/J$ from $0$ (blue) to $10$ (red).}
	\label{fig:3}
\end{figure}

\subsection{Moir\'e patterns and associated density of states\label{subsec:band}}

As it occurs with the honeycomb geometry~\cite{dossantos07a,shallcross10a,trambly10a,morrel10a,bistritzer11a}, only certain angles give rise to a crystal like configuration with a new periodicity (Moir\'e patterns). For square geometries the conmensuration angles can be written as:
\begin{align}
\label{eq:angle}
\theta(m,n)=\arccos\left(\frac{2 m n}{m^2+n^2}\right),n,m\in\mathbb{Z}\,,
\end{align}
where we can restrict to $\theta(m,n)<\pi/4$ due to the symmetry of the layers.

Since the band structure and density of states are properties of the non-interacting part of the Hamiltonian we only need to consider $H_a+H_b+H_{ab}$ for the calculation. Furthermore, in this manuscript we restrict to the strong confinement limit ($L_0\ll \lambda/2$) such that we only include nearest neighbour hopping for the intralayer hopping with strength $J$, and on-site inter-layer hopping, $J_{\perp}(\RR)=J_\perp \delta_{\RR,0}$. For the large angles considered along this manuscript this minimal model will be a good approximation with small quantitative differences when longer range hoppings are included. 

 To calculate the associated density of states we define periodic boundary conditions, and diagonalize the Hamiltonian in momentum space for a discretization of $\kk$ space of $N\times N$ points. The density of states is then given by:
\begin{align}
D(\omega)\propto \sum_{j=1}^{N_\mathrm{M}}\sum_{\kk} \mathcal{N}(\omega-\Delta\omega<\omega_j(\kk)<\omega)\,,
\end{align}
where $N_\mathrm{M}=2(n^2+m^2)$ is the total number of $a/b$ lattice sites per unit cell (which is also the number of bands $\omega_i(\kk)$), and $\mathcal{N}(x)$ is a function that counts the number of states in a given frequency interval of width $\Delta\omega$, that we set to be $2\pi/N$ for the figures of this manuscript.

In Fig.~\ref{fig:2}(a) we plot the density of states at a fixed angle $\theta(2,1)\approx 36.9^\circ$, whose Moir\'e pattern is plotted in Fig.~\ref{fig:2}(b), for several $J_\perp/J\in [0,4]$. We observe several phenomena: i) the central van-Hove singularity splits for increasing $J_\perp/J$; ii) there appear extra divergences in the middle of the band; iii) at a critical $J_\perp/J\approx 1.7$ (for this angle) two of them separate as isolated bands in the lower/upper band part of the spectrum. These isolated bands get flatter as the angle decreases. This is clear in Fig.~\ref{fig:2}(c) where we zoom around these bands for fixed $J_\perp/J=4$, and plot the density of states  for several angles ranging from $36.9^\circ$ to $12.1^\circ$. Though not shown, the critical ratio $J_\perp/J$ where these isolated upper and lower bands split from the rest also decreases with smaller angles. Thus the possibility to tune $J_\perp/J$ provide us a knob to observe narrow bands or the splitting van-Hove singularities with larger rotation angles. 

To understand better the origin of the different contributions appearing in the density of states, we calculate the projected band-structure for $J_\perp/J=4$ and $\theta(2,1)$ along the square symmetry points, and plot it together with the associated density of states in Fig.~\ref{fig:2}(d). There, we can identify that the band-structure of the narrow upper and lower bands is similar to a square tight-binding model, e.g., with a saddle-point at the $X$ point, but with reduced band-width. As it happens in twisted bilayer graphene~\cite{ni15a,sunku18a}, the band-width is reduced because the hopping occurs mostly along the diagonals of the new primitive cell.  Even more interesting is what happens in the middle of the band, where the accidental degeneracy between several bands give rise to different type of van-Hove singularities. For example, in $\omega\approx J (2.2J)$, six (two) bands touch at the $M (X)$-point leading to an asymmetric van-Hove singularity in the density of states. Similar asymmetric singularities have also been reported in twisted bilayer graphene~\cite{kerelsky18a}, and explained in terms of high-order van-Hove singularities~\cite{yuan19a}.

\subsection{Considerations about other geometries~\label{subsec:dirac}}

Along this manuscript we have focused on square-like potentials whose physics is dominated by the Van-Hove singularities explored in Fig.~\ref{fig:2}. For completeness, we want to note that it is possible to adapt some of the schemes discussed in \ref{sec:twist}, e.g., the one using turnout wavelengths, together with existing experimental ideas~\cite{tarruell12a,soltan11a} to simulate Dirac like physics. The turnout wavelength scheme is convenient because it does not rely on having orthogonal polarization for the twisted standing waves, but rather on different frequencies (and angled beams), providing more freedom to choose the polarization for the rotated and unrotated potentials without having to worry about cross interference between them.

One possibility to obtain Dirac-like physics consists in keeping the square geometry behaviour but with staggered hoppings, forming the so-called brick-wall lattice. This configuration has already been realized experimentally for a single atomic state in Ref.~\cite{tarruell12a}. For each layer it requires two retro-flected lasers in the X and Y directions with out-of-plane linear polarization and the same frequency, plus an additional retro-reflected laser in the X direction with slight different frequency such that its contribution adds up incoherently to the potential of the first two. With an appropriate choice of intensities (see Ref.~\cite{tarruell12a}), this potential gives rise to an energy dispersion with two Dirac points. However, since the symmetry of the layer is different than that of graphene, the underlying physics emerging under rotation will be different.

Another possibility consists in directly creating honeycomb lattices, as experimentally done in Ref.~\cite{soltan11a} for a single atomic state. This can be done by sending three counter-propagating lasers within the same plane with $60^\circ$ angles between them and linear polarization for each of the turnout wavelengths, using angled beams for the one with larger momentum. In this case, the Moir\'e patterns occur for the same angles than in graphene~\cite{trambly10a}:
\begin{align}
\theta_\mathrm{hc}(m,n)=\arccos\left(\frac{n^2+4 n m+m^2}{2(m^2+n^2+nm)}\right),n,m\in\mathbb{Z}\,,
\end{align}

Since one of the features of this platform is the possibility to tune the inter/intra-layer hopping ratio, in Fig.~\ref{fig:3} we calculate the change in the density of states (panel a), and band structure (panel b) for several $J_\perp/J$ and a honeycomb bilayer with fixed angle $\theta_\mathrm{hc}(2,1)$ (zooming around the Dirac point present when $J_\perp/J=0$). As in the square lattice model, we use a minimal hopping model restricting to nearest neighbour intra-layer hoppings and local inter-layer ones. Despite the simplicity of the model, we already observe how increasing the interlayer hopping can also be used to shift Van-Hove singularities close to Dirac points and to narrow the bands around the Dirac point for a given angle.

\section{Probing twisted bilayer systems by spontaneous emission~\label{sec:qo}}

Finally, to illustrate the potential of the proposed setup to explore physics beyond the one observed in solid state platforms, we place it in the context of the recent experiments simulating quantum optical phenomena with matter-waves~\cite{krinner18a}. There, one studies the physics of one or several (simulated) emitters interacting with the matter-waves propagating in an structured optical potential. As shown in recent works~\cite{douglas15a,gonzaleztudela15c,gonzaleztudela17a,gonzaleztudela18c,gonzaleztudela18f,gonzaleztudela18d}, these structured "photons" can give rise to unconventional emitter dynamics and interactions with no analogue in standard photonic environments, and which can be instrumental, for example, for quantum simulation of long-range interacting spin systems~\cite{douglas15a,gonzaleztudela15c} or quantum chemistry problems~\cite{arguello18a}.

Apart from the states playing the role of matter-waves (the $a/b$ levels), in these experiments one requires an additional long-lived atomic state, $c$, to play the role of the emitter. Depending on the realization considered in section~\ref{subsec:alkali}, this can be either the ground state $^1 S_0$ (for the ones relying in orthogonally polarized standing waves), or the other metastable excited state $^3 P_2$ for the one exploiting turnout wavelengths. The advantage is that the potential for this state does not need to have the same periodicity than $V_{a,b}(\RR)$, such that the laser can have a different frequency. Another possibility is to use optical tweezers to trap them at controlled positions instead of retro-reflected lasers. In all cases, the potential depth must be deep enough such that once the atom is in the $c$ state in a given position it can not hop anywhere, but only emit into matter-waves. The Hamiltonian of these emitter-like atoms can be written as:
\begin{align}
H_c=\omega_c\sum_{i}c_i^\dagger c_i+\frac{U_c}{2} \sum_{i} c_i^\dagger c_i (c_i^\dagger c_i-1)\,,
\end{align}
where $U_c$ is the on-site interaction of this atomic state. As explained in section~\ref{sec:band}, these interactions can be controlled, such that one can tune between having bosonic emitters when $U_c=0$ to purely two level ones $c_i\approx \sigma_i$ when $U_C\rightarrow \infty$.

The coupling between this internal atomic state and the photonic modes can be done in a similar way than the inter-layer hopping one (see Eqs.~\ref{eq:Hab},\ref{eq:inter}): either via a two-photon Raman transition through an auxiliary state, or with a direct microwave/optical transition depending on the level structure considered. Assuming we couple the $c$ state only to the $a$-modes, the simulated light-matter Hamiltonian in the strong confinement limit can be written as:
\begin{align}
H_\intt=g\sum_{i}\left(a_{\nn_i}^\dagger c_i+\mathrm{H.c}\right) \,,
\end{align}
where $\nn_i$ denotes the lattice position where the $i$-th emitter-like atom couples to, and $g$ is the coupling strength which we assume to be real. For more details we refer the reader to the original Refs.~\cite{devega08a,navarretebenlloch11a} where all the details of this type of simulation were laid out.

To illustrate the physics that can emerge in these bilayer reservoirs we concentrate in the spontaneous emission dynamics of a single emitter. This means we assume to have a single emitter-like atom that is initially excited, $\ket{\Psi(0)}=c^\dagger\ket{\mathrm{vac}}$, and study its evolution under the global light-matter Hamiltonian given by: $H=H_c+H_\intt+H_B$ as a function of time. Since $H$ conserves the number of excitations, the wavefunction at any time $\ket{\Psi(t)}=e^{-i Ht}\ket{\Psi(0)}$ reads:
\begin{align}
\ket{\Psi(t)}=\left(C_e(t)c^\dagger+\sum_{\alpha,\RR} C_{\alpha,\nn}(t)\alpha_\nn^\dagger\right) \ket{\mathrm{vac}}\,,
\end{align}

Notice that since we have a single excitation in the system at any time, the interactions $U_\alpha$ or the statistics of the atoms play no role in the dynamics, which only depends on the system-bath coupling $g$, the inter/intra-layer coupling $J_\perp,J$, and the detuning between the emitter-like and the bath frequencies $\Delta=\omega_c-\omega_t$.  Perturbative treatments, like Markov approximation~\cite{cohenbook92a}, predict an exponential decay of the population, i.e.,  $|C_e(t)|^2\approx e^{-\Gamma_M t}$, with $\Gamma_M\propto g^2 D(\omega_c)$. However, since the density of states of these systems has a non-trivial structure, as shown in Fig.~\ref{fig:2}, the dynamics can differ significantly from this prediction.

\begin{figure}[tb]
	\centering
	\includegraphics[width=0.49\textwidth]{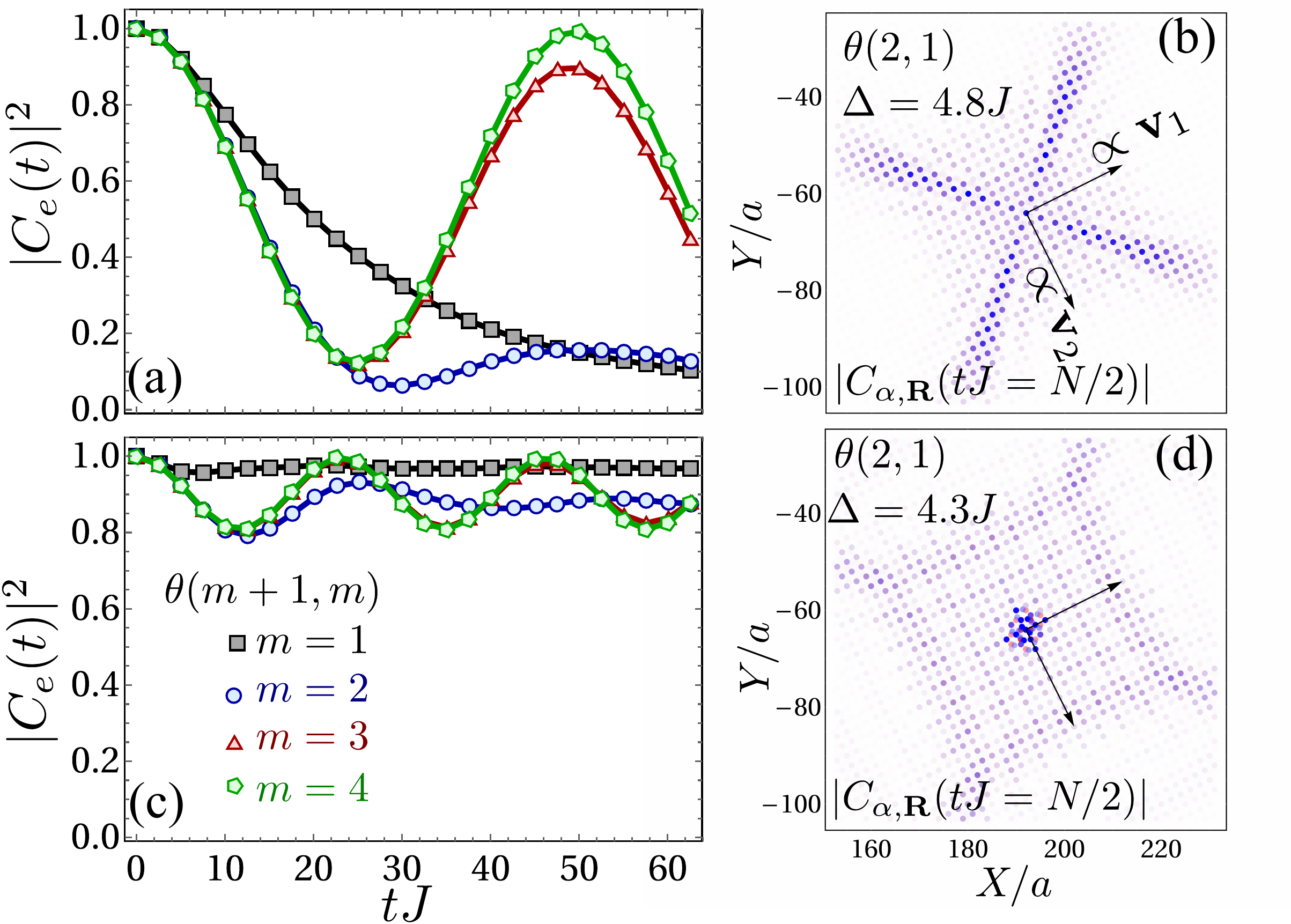}
	\caption{Spontaneous emission of matter-waves from a simulated emitter coupled with strength $g=0.1J$ to a bilayer bath with size $128\times 128$ and $J_\perp/J=4$. (a) [(c)] Excited state population for several angles as depicted in the legend and $\Delta$ chosen to be in the middle (out of) of the upper isolated band, i.e., $\Delta=4.8$ [$4.3$] for $m=1$ and $\Delta=5$ [$4.8$] for the rest. (b) [(d)] Associated bath probability amplitude in the A/B bilayers (in red/blue respectively) after a time $tJ=N/2$ for a emitter with energy $\Delta/J=4.8$ [$4.3$], and a twisting angle $\theta(2,1)$.}
	\label{fig:4}
\end{figure}

To exemplify it, in Figs.~\ref{fig:4}(a-b) we plot the exact spontaneous emission dynamics of a single emitter coupled locally with $g=0.1J$ to a twisted square bilayer photonic reservoir with $J_\perp/J=4$. We consider two different situations: when the emitter's frequency exactly matches the peak of the upper separated band (that we call the resonant case) in Figs.~\ref{fig:4}(a-b), and when the atomic frequency lies in the bandgap but close to lower edge of this separated band (that we call  the off-resonant case) in Figs~\ref{fig:4}(c-d). The exact parameters are depicted in the legend and caption of Fig.~\ref{fig:4}. For the resonant configuration (see Fig.~\ref{fig:4}(a)) we observe that for the larger angle $\theta(2,1)$ the dynamics is mostly irreversible, following approximately an exponential decay. However, as the angle decreases the dynamics becomes more and more reversible (non-Markovian). The underlying reason is that the effective bath tunnelling rate is being reduced with the twisting angle and starts being comparable with $g$. For very small angles, the emitter exchanges excitations mostly with the localized mode that emerges in the Moir\'e supercell, leading the dynamics observed in the figure. This transition between Markovian and non-Markovian dynamics also occurs when the emitter's frequency lies in the band-gap, shown in Fig.~\ref{fig:4}(c). In that case, the emitter goes from the no-decay situation predicted by Markov to a fractional decay one~\cite{john94a} in which the emitter decays partially and oscillates around a constant value. Thus, the twisting angle can be used as a knob to boost non-Markovian behaviour in such systems without increasing system-bath couplings.

Another interesting feature to consider is the behaviour of the matter-waves created in both the resonant and off-resonant situations, plotted in Figs.~\ref{fig:4}(b) and (d), respectively. When the emitter is resonant with the middle-band peak, the matter-wave pattern is highly anisotropic emitting mostly in $2$ directions which correspond to the diagonal directions of the emergent Moir\'e primitive cell. In the band-gap case, the bath excitations become exponentially localized around the emitter, forming a bound-state around them~\cite{bykov75a,john90a,kurizki90a}. However, this bound state is anisotropic, being more extended again in the diagonal direction of the new primitive cell. We believe that the preferential emission (or localization) in the (off-)resonant cases along these new Moir\'e directions has a similar origin than the coupled-wire picture~\cite{wu19a} used to explain the results in twisted bilayer graphene, or the response of the material when it is probed by local plasmons~\cite{ni15a,sunku18a} or preparing a localized electron wavepacket~\cite{do19a}. From the quantum optical perspective, the main interest of such non-trivial matter-wave behaviour is that when putting more than one emitter, the induced collective decays ($\gamma_{ij}$) and dipole couplings ($J_{ij}$) will follow the same patterns~\cite{dicke54a,lehmberg70a,lehmberg70b} giving rise to very exotic equilibrium and out-of equilibrium many-body dynamics. For example, when the emitter's frequency lie in the band-gap, the dynamics will be fully coherent and governed only by the dipole-dipole couplings:
\begin{align}
H_\mathrm{spin}=\sum_{i,j}J_{i,j}c_i^\dagger c_j\,,
\end{align}
which in the blockaded limit ($U_c\rightarrow \infty$) corresponds to an effective spin model with Moir\'e-like interactions.

\section{Conclusions and outlook~\label{sec:conclusions}}

Summing up, we have shown how to create two rotated state-dependent optical potentials to simulate the physics of twisted bilayers using cold-atoms. The possibility of tuning independently the inter/intra-layer hopping ratio opens the door to observe narrow bands or to move and split van-Hove singularities for larger rotation angles than in the solid state realizations. The potential of the proposed setup is three-fold: first, since the simulated particles interact locally and do not couple to phonons this setup can be used to discard some of the explanations introduced to explain the strongly correlated phenomena observed in experiments. Second, the possibility of observing similar physics with larger rotation angles simplifies ab-initio calculation and can be used as a platform to benchmark some of the effective descriptions made for the small rotation angles. Finally, as we have shown in the last part of the manuscript, the setup has the potential to explore different physics from solid-state platforms, e.g., quantum optical phenomena with structured Moir\'e photonic baths. To illustrate it,  we have studied the spontaneous decay of a single emitter coupled to such type of bath, and predicted how the twisting angle can be used as a knob to enhance non-Markovianity, or to induce non-trivial emission patters and emitter-emitter couplings which follow the emergent Moir\'e geometry.

Taking this work as a basis, we believe there are still many open questions that will trigger more studies on the subject. For example, from the implementation point of view it will be interesting to find simpler atomic/laser schemes for the simulation of Dirac like physics as well as other ones that allow the simulation multi-layered materials that also show flat-bands. On the fundamental side, there are several relevant questions to answer such as the role of longer range hoppings when we tune away from the strong confinement situation, or whether we can observe the magic-angle behaviour (with a complete flattening of the band-structure) for larger angles by increasing $J_\perp/J$. From the perspective of the quantum optical experiments of matter-waves, it will be interesting to explore what kind of dynamics emerges from the higher-order van-Hove singularities that we found in the square twisted bilayers, and which also appear in the graphene case~\cite{yuan19a}.

\section*{Acknowledgements}

AGT acknowledges support from CSIC Research Platform PTI-001 and from the national project PGC2018-094792-B-I00 from Ministerio de Ciencia e Innovacion. JIC acknowledges the ERC Advanced Grant QENOCOBA under the EU Horizon 2020 program (grant agreement 742102). The authors acknowledge very useful discussions with S.~Blatt, S.~F\"olling, I.~Bloch, J.~Kn\"orzer, T.~Udem, G. Giedke, and J.~Gonz\'alez.


\bibliography{Sci,books}

\end{document}